\documentclass[twoside,11pt]{article}

\usepackage{jmlr2e}

\usepackage[latin2]{inputenc}
\usepackage[T1]{fontenc}
\usepackage{paralist,booktabs,amsmath,bm,alltt}
\usepackage[mathscr]{eucal}
\usepackage[colorlinks=false,allbordercolors={1 1 1}]{hyperref}

\renewcommand{\b}{\mathbf}
\newcommand{\R}{\mathbb{R}}

\jmlrheading{15}{2014}{283-287}{10/12; Revised 9/13}{1/14}{Zoltán Szabó}

\ShortHeadings{Information Theoretical Estimators (ITE) Toolbox}{Zoltán Szabó}
\firstpageno{283}

\begin{document}

\title{Information Theoretical Estimators Toolbox}

\author{\name Zoltán Szabó\thanks{The authors would like to thank the anonymous reviewers for their valuable suggestions. This work was supported by the Gatsby Charitable Foundation.
The research was carried out as part of the EITKIC\_12-1-2012-0001 project, which is supported by the Hungarian Government, managed by the National Development
Agency, financed by the Research and Technology Innovation Fund and was performed in cooperation with the EIT ICT Labs Budapest Associate Partner Group.
(\url{www.ictlabs.elte.hu}). The Project was supported by the European Union and co-financed by the European Social Fund (grant agreement no.\ T\'AMOP
4.2.1/B-09/1/KMR-2010-0003).} \email zoltan.szabo@gatsby.ucl.ac.uk \\
      \addr Gatsby Computational Neuroscience Unit\\
	    Centre for Computational Statistics and Machine Learning\\
	    University College London\\
	    Alexandra House, 17 Queen Square, London - WC1N 3AR}

\editor{Balázs Kégl}

\maketitle

\begin{abstract}
We present ITE (information theoretical estimators) a free and open source, multi-platform, Matlab/Octave toolbox that is capable of estimating many different variants of 
entropy, mutual information, divergence, association measures, cross quantities, and kernels on distributions.
Thanks to its highly modular design, ITE supports additionally (i) the combinations of the estimation techniques, (ii) the easy
construction and embedding of novel information theoretical estimators, and (iii) their immediate application in information theoretical optimization problems.
ITE also includes a prototype application in a central problem class of signal processing, independent subspace analysis and its extensions. 
\end{abstract}


\section{Introduction}
Since the pioneering work of \citet{shannon48mathematical}, \emph{entropy}, \emph{mutual information},\footnote{The Shannon mutual information is also 
known in the literature as the special case of total correlation or multi-information when two variables are considered.} \emph{association}, \emph{divergence} measures and \emph{kernels on distributions} have found a broad range of applications in many areas of machine learning \citep{beirlant97nonparametric,wang09divergence,villmann10mathematical,basseville12divergence,poczos12copula,sriperumbudur12empirical}.
Entropies provide a natural notion to quantify the \emph{uncertainty} of random variables, mutual information and association indices
measure the \emph{dependence} among its arguments, divergences and kernels offer efficient tools to define the `distance' and the inner product
of probability measures, respectively. 

A central problem based on information theoretical objectives in signal processing is independent subspace
analysis (ISA; \citealt{cardoso98multidimensional}), a cocktail party problem with \emph{independent groups}. 
One of the most relevant and fundamental hypotheses of the ISA research is the \emph{ISA separation principle} \citep{cardoso98multidimensional}: the ISA task can be solved by ICA (ISA with one-dimensional sources, \citealt{hyvarinen01independent,cichocki02adaptive,choi05blind}) followed by clustering of the ICA elements.
This principle (i) forms the basis of the state-of-the-art ISA algorithms, (ii) can be used to design algorithms that scale well and
efficiently estimate the dimensions of the hidden sources, (iii) has been recently proved \citep{szabo07undercomplete_TCC}
and (iv) can be extended to different \mbox{linear-,}
controlled-, \mbox{post nonlinear-,} complex valued-, partially observed systems, as well as to systems
with nonparametric source dynamics. For a recent review on the topic and ISA applications, see \citet{szabo12separation}.

Although there exist many exciting applications of information theoretical measures,
to the best of our knowledge, available packages in this domain focus on (i) discrete variables, or
(ii) quite specialized applications/information theoretical estimation methods.
Our \textbf{goal} is to fill this serious gap by coming up with a (i) highly modular, (ii) free and open source, (iii) multi-platform toolbox, the
ITE (information theoretical estimators) package, which focuses on \emph{continuous} variables and 
\begin{compactenum}
  \item is capable of estimating \emph{many} different kind of entropy, mutual information, association, divergence measures, distribution kernels based on
  nonparametric methods.\footnote{It is highly advantageous to apply nonparametric approaches: the `opposite' plug-in type methods---estimating the underlying densities---scale poorly as the dimension is increasing.}
  \item offers a \emph{simple and unified framework} to
	(i) easily construct new estimators from existing ones or from scratch, and
	(ii) transparently use the obtained estimators in information theoretical optimization problems,
  \item with a \emph{prototype application} in ISA and its extensions.
\end{compactenum}

\section{Library Overview}\label{sec:H-I-D}
Below we provide a brief overview of the ITE package:
\begin{compactdesc}
 \item[Information Theoretical Measures:]
    The ITE toolbox is capable of estimating numerous important information theoretical quantities including
    \begin{compactdesc}
	\item[Entropy:] Shannon-, Rényi-, Tsallis-, complex-, $\Phi$-, Sharma-Mittal entropy,  
	\item[Mutual information:]
	      generalized variance, kernel canonical correlation analysis, kernel generalized variance, Hilbert-Schmidt independence criterion, 
	      Shannon-, $L_2$-, Rényi-, Tsallis-, Cauchy-Schwartz quadratic-, Euclidean distance based \mbox{quadratic-,} complex-,  $\chi^2$ mutual information; copula-based kernel dependency, multivariate version of Hoeffding's $\Phi$, 
	      Schweizer-Wolff's $\sigma$ and $\kappa$,  distance covariance and correlation, approximate correntropy independence measure,
	\item[Divergence:]
	    Kullback-Leibler-, $L_2$-, Rényi-, Tsallis-, Cauchy-Schwartz-, Euclidean distance based-, Jensen-Shannon-, Jensen-Rényi-, Jensen-Tsallis-, K-, L-, Pearson $\chi^2$-, f-divergences; 
	    Hellinger-, Bhattacharyya-, energy-, (non-)symmetric Bregman-, J-distance; maximum mean discrepancy, 
	\item[Association measure:] multivariate (conditional) extensions of Spearman's $\rho$, (centered) correntropy, correntropy induced metric, correntropy coefficient, 
	centered correntropy induced metric, multivariate extension of Blomqvist's $\beta$, lower and upper tail dependence via conditional Spearman's $\rho$,
	\item[Cross quantity:] cross-entropy,
	\item[Distribution kernel:] expected-, Bhattacharyya-, probability product-, (exponentiated) Jensen-Shannon-, (exponentiated) Jensen-Tsallis-, exponentiated Jensen-R{\'e}nyi kernel.
    \end{compactdesc}
  \item[Independent Process Analysis (IPA):] 
      ITE offers solution methods for independent subspace analysis (ISA) and its extensions to different \mbox{linear-}, controlled-, post nonlinear-, complex valued-, partially observed systems, as well as to systems
      with nonparametric source dynamics; combinations are also possible. The solutions are based on 
      the ISA separation principle and its generalizations \citep{szabo12separation}.
  \item[Quick Tests:] Beyond IPA, ITE provides quick tests to study the efficiency of the estimators. These tests cover (i) analytical value vs.\ estimation, (ii) positive semi-definiteness of Gram matrices defined by distribution kernels and (iii) image registration problems.
  \item[Modularity:] The core idea behind the design of ITE is modularity. The modularity is based on the following four pillars:
      \begin{compactenum}
	  \item 
	      The estimation of many information theoretical quantities can be reduced to 
	       k-nearest neighbor-, minimum spanning tree computation, random projection, ensemble technique, 
	      copula estimation, kernel methods. 
	  \item The ISA separation principle and its extensions make it possible to decompose the solutions of the 
	      IPA problem family to ICA, clustering, ISA, AR (autoregressive)-, ARX- (AR with exogenous input) and mAR (AR with missing values) identification, gaussianization and nonparametric regression subtasks.
	  \item Information theoretical identities can relate numerous entropy, mutual information, association, cross- and divergence measures, distribution kernels \citep{cover91elements}.
	  \item ISA can be formulated via information theoretical objectives \citep{szabo07undercomplete_TCC}:
	\begin{align*}
	  J_{\text{I}}(\b{P}) &= I\left(\b{y}^1,\ldots,\b{y}^M\right),\hspace*{0.75cm} 
	  J_{\text{Irecursive}}(\b{P}) = \sum_{m=1}^{M-1} I\left(\b{y}^m,\left[\b{y}^{m+1},...,\b{y}^M\right]\right),\\
	  J_{\text{sumH}}(\b{P}) &= \sum_{m=1}^M H\left(\b{y}^m\right),\hspace*{1.8cm}
	  J_{\text{sum-I}}(\b{P}) = -\sum_{m=1}^M I\left(y_1^m,...,y_{d_m}^M\right),\\
	  J_{\text{Ipairwise}}(\b{P}) &= \hspace*{-0.2cm}\sum_{m_1\ne m_2} I\left(\b{y}^{m_1},\b{y}^{m_2}\right),\hspace*{0.05cm}
	  J_{\text{Ipairwise1d}}(\b{P}) = \hspace*{-0.2cm}\sum_{m_1\ne m_2}^M\sum_{i_1=1}^{d_{m_1}}\sum_{i_2=1}^{d_{m_2}} I\left(y_{i_1}^{m_1},y_{i_2}^{m_2}\right),
	\end{align*}
	where the minimizations are w.r.t.\ the optimal clustering ($\b{P}$) of the ICA elements.
      \end{compactenum}
  \item[Dedicated Subtask Solvers, Extension:] The ITE package offers dedicated solvers for the obtained subproblems 
    detailed in `\emph{Modularity}:1-2'. Thanks to this flexibility, extension of ITE can be done effortlessly: it is sufficient to add a new switch entry 
    in the subtask solver.
  \item[Base and Meta Estimators:] One can \emph{derive} new, \emph{meta} (entropy, mutual information, association, divergence, cross quantity, distribution kernel) estimators 
     in ITE from existing base or meta ones by 
    `\emph{Modularity}:3'. The calling syntax of base and meta methods are completely identical thanks to the underlying unified template
    structure followed by the estimators. The ITE package also supports an indicator for the importance of multiplicative 
    constants.\footnote{In many applications, it is completely irrelevant whether we estimate, for example, $H(\b{y})$ or $cH(\b{y})$, where $c=c(d)$ is a constant depending only on the \emph{dimension} of $\b{y}\in\R^d$ ($d$), but \emph{not on the distribution} of $\b{y}$. 
    Such `up to proportional factor' estimations can often be carried out more efficiently.}
  
    We illustrate how easily one can estimate information theoretical quantities in ITE:
    \begin{alltt}
  >Y1 = rand(3,1000); Y2 = rand(3,2000);   %data of interest
  >mult = 1;                   %multiplicative constant is important
  >co = D_initialization('Jdistance',mult);%initialize the estimator
  >D = D_estimation(Y1,Y2,co);             %estimation
    \end{alltt}\vspace*{-0.4cm}
    Next, we demonstrate how one can construct meta estimators in ITE. 
    We consider the definitions of the initialization and the estimation of the J-distance. 
    The KL-divergence, which is symmetrised in J-distance, is estimated based on the existing k-nearest neighbor technique.
    \begin{alltt}
  function [co] = DJdistance_initialization(mult)
  co.name = 'Jdistance';           %name of the estimator
  co.mult = mult;                  %importance of multiplicative const.
  co.member_name = 'KL_kNN_k';     %method used for KL estimation
  co.member_co = D_initialization(co.member_name,mult); %initialization
\end{alltt}\vspace*{0.2cm}
\begin{alltt}
  function [D_J] = DJdistance_estimation(X,Y,co)
  D_J = D_estimation(X,Y,co.member_co) + D_estimation(Y,X,co.member_co);
    \end{alltt}\vspace*{-0.4cm}
\item[ISA Objectives and Optimization:] Due to the unified syntax of the estimators, 
    one can formulate and solve information theoretical optimization problems in ITE in a high-level view. 
    Our example included in ITE is ISA (and its extensions) whose objective can be expressed 
    by entropy and mutual information terms, see `\emph{Modularity}:4'. The unified
    template structure in ITE makes it possible to use \emph{any} of the estimators (base/meta) in these cost functions.

    A further attractive aspect of ITE is that even in case of 
    unknown subspace dimensions, it offers well-scaling approximation schemes based on spectral clustering methods. Such methods
    are (i) robust and (ii) scale excellently, a single general desktop computer can handle about a million observations---in our case estimated ICA elements---within several minutes \citep{yan09fast}. 
\end{compactdesc}

\section{Availability and Requirements}
The ITE package is \emph{self-contained}, it only needs a Matlab or an Octave environment\footnote{See \url{http://www.mathworks.com/products/matlab/} and \url{http://www.gnu.org/software/octave/}.} with standard toolboxes.
ITE is \emph{multi-platform}, it has been extensively tested on Windows and Linux; since it is made of standard Matlab/Octave and C++ files, it is expected to work on alternative platforms as well.\footnote{On Windows (Linux) we suggest using the Visual C++ (GCC) compiler.}
ITE is released under the free and open source GNU GPLv3 ($\ge$) license. 
The accompanying source code and the documentation of the toolbox has been enriched with numerous comments, examples, detailed instructions for extensions, and pointers where the interested user can find further mathematical details about the embodied techniques.
The ITE package is available at \url{https://bitbucket.org/szzoli/ite/}.


\bibliography{szabo14a}

\end{document}